\documentclass[aps,prb,reprint,groupedaddress]{revtex4-1}

\usepackage{hyperref}
\usepackage{graphicx}
\usepackage{multirow}

\begin{document}


\title{Electronic transport properties of nitrate-doped carbon nanotube networks}


\author{T. Ketolainen}
\email[]{tomi.ketolainen@aalto.fi}
\author{V. Havu}
\author{E. \"{O}. J\'{o}nsson}
\author{M. J. Puska}
\affiliation{COMP, Department of Applied Physics, Aalto University, P.O. Box 11100, FI-00076 Aalto, Finland}


\date{\today}

\begin{abstract}
The conductivity of carbon nanotube (CNT) networks can be improved markedly by doping with nitric acid. In the present work, CNTs and junctions of CNTs functionalized with NO$_3$ molecules are investigated to understand the microscopic mechanism of nitric acid doping. According to our density functional theory band structure calculations, there is charge transfer from the CNT to adsorbed molecules indicating p-type doping. The average doping efficiency of the NO$_3$ molecules is higher if the NO$_3$ molecules form complexes with water molecules. In addition to electron transport along individual CNTs, we have also studied electron transport between different types (metallic, semiconducting) of CNTs. Reflecting the differences in the electronic structures of semiconducting and metallic CNTs, we have found that besides turning semiconducting CNTs metallic, doping further increases electron transport most efficiently along semiconducting CNTs as well as through a junction between them.
\end{abstract}


\maketitle


\section{Introduction\label{sec:introduction}}

Flexible carbon nanotube (CNT) thin films are materials, where randomly oriented CNTs form a network structure \cite{Yu2016}. The fabrication methods of CNT thin films have improved remarkably and several novel applications of the CNT networks have been introduced. It may be even possible to replace the commonly used indium tin oxide films by CNT thin films. Advantages of the CNT thin films are their low absorption in a broad range of optical wavelengths \cite{Hu2009} and a possibility to bend the films without lowering the conductivity significantly \cite{Wang2012}.

In applications, CNT thin films with high electrical conductivity and optical transparency are required. For this purpose, a significant enhancement of the conductivity of the CNT thin films has been found, for example, in the case of AuCl$_3$ doping \cite{Kim2008}. Computational studies have revealed that formation of AuCl$_4$ molecules or anions on semiconducting CNTs leads to a p-type doping effect, i.e. to electron transfer to acceptor molecules leaving behind holes in the substrate CNTs \cite{Murat2014}. Furthermore, in our recent study we have found a significant enhancement of the electron transport through a junction of semiconducting CNTs due to the AuCl$_4$ doping and the improvement has been shown to be robust with respect to the molecular concentration provided that it is large enough \cite{Ketolainen2017}. Recently, iodine monochloride and iodine monobromide have also been used to dope CNT thin films, indicating a remarkable reduction of their electrical resistivity \cite{Janas2017}. Another example of improving the conductivity of CNT thin films is doping them with I chains or CuCl as presented experimentally in Ref.~\cite{Tsebro2016}. Moreover, previous computational studies of CNT junctions with transition metal linker atoms have shown enhanced junction conductances \cite{Ketolainen2015}.

A common method to improve the conductivity of CNT thin films is to dope them with nitric acid (see, for example, Refs.~\cite{Graupner2003}, \cite{Zhou2005}, and \cite{Kaskela2010}). Experimentally, the nitric-acid doping has been shown to reduce the junction resistances as well as increase the conductivity of individual CNTs in the network \cite{Znidarsic2013,Jeong2015}. However, the mechanism behind the nitric-acid doping of CNTs is not well-known. It is assumed to be related to nitrogen oxide molecules bound to CNTs. The adsorption of various NO$_x$ molecules on CNTs has been studied using density functional theory (DFT) in order to understand their detection in gas sensing applications \cite{Peng2004}. Computationally, it has also been found that NO$_2$ molecules can form NO$_3$ molecules and the binding energy is higher in the case of molecular pairs \cite{Dai2009}. According to previous experimental and computational studies, the strength and type of the adsorption of nitrogen oxides are also affected by the metallicity of the CNT \cite{Ruiz-Soria2014}. In addition to the chirality, recent DFT calculations of the nitric-acid treated CNTs have shown that the charge state of the CNT also influences the adsorption process \cite{Kanai2010}.

The electronic structure and transport properties of CNT junctions need to be considered in more detail to understand the effect of nitric acid on the conductivity of CNT networks. Earlier experimental studies have shown that the resistance of the CNT junction depends largely on the chiralities of the CNTs and the resistance may be especially high when a metallic and a semiconducting CNT form a junction with a Schottky barrier \cite{Fuhrer2000}. Furthermore, CNT junctions with physisorbed O$_2$ and N$_2$ molecules have been investigated previously experimentally and theoretically \cite{Mowbray2009}. According to the computational part of that work, molecules physisorbed in the vicinity of the junctions lead to an improvement of the conductance of the junction, which has been explained to result from improved tunneling probability due to hopping via molecular orbitals \cite{Mowbray2009}.

In this work, we consider the influence of nitric acid on CNT networks by studying CNTs with adsorbed NO$_3$ molecules and junctions of NO$_3$-doped CNTs. On the basis of simple electron counting, NO$_3$ molecules are expected to work as acceptors (or anions). More specifically, optimal geometries, band structures, and electronic transmission functions for NO$_3$-doped CNTs are determined using DFT. The DFT calculations reveal that the physisorbed NO$_3$ molecules receive electrons from the CNT resulting in a downshift of the CNT Fermi level and ensuing p-type doping of the CNT. Intratube as well as intertube conductances achieved in doping depend on the chiralities of the CNTs. Semiconducting CNTs become conductive and the electron transport can be further efficiently increased by lowering the Fermi level below van Hove singularities when they are doped with NO$_3$ molecules. Moreover, junctions of two NO$_3$-doped semiconducting CNTs with a large number of molecules show a clearly improved conductance so that the improvement can be a factor of ten or even more. The doping of metallic CNTs lowers the intratube conductances a little depending on the molecular concentration. However, the doping does not lower the Fermi level as efficiently toward the van Hove singularities of the CNT electronic structure as in the case of semiconducting CNTs. This diminishes the effect of doping on the intratube and intertube conductances. The clearly dissimilar electronic structures of semiconducting and metallic CNTs hinder to enhance the conductance of their junctions by doping. Similarly to the predictions for complex molecular systems \cite{Jonsson2011}, we have found that water molecules coordinated to NO$_3$ molecules enhance the doping effect both in semiconducting and in metallic CNTs.

The structure of this article is as follows. The main aspects of the methods and investigated systems are presented in Sec.~\ref{sec:methods_and_systems}. Then the geometries and band structures with the ensuing doping efficiencies are discussed at the beginning of Sec.~\ref{sec:results}. The essential topics at the end of this section are electron transport in NO$_3$-doped CNTs and CNT junctions. A brief summary of the results is given in Sec.~\ref{sec:conclusions}.

\section{Methodology and systems\label{sec:methods_and_systems}}

The calculations presented are carried out using two different electronic-structure codes, FHI-aims \cite{Blum2009,Levchenko2015,Ihrig2015} and GPAW \cite{Enkovaara2010,Mortensen2005}. We use the FHI-aims code to perform most of the calculations apart from calculations including the Perdew-Zunger self-interaction correction (SIC) \cite{Perdew1981}. The latter calculations are carried out with the GPAW electronic-structure code package. The FHI-aims code package is based on numeric atom-centered orbitals and also provides tools for studying transport properties of materials. Two exchange-correlation functionals are used, namely the generalized-gradient approximation functional PBE \cite{Perdew1996} and the hybrid functional HSE06 \cite{Krukau2006}. Since van der Waals interactions can be significant in nanocarbon systems, the Tkatchenko-Scheffler van der Waals correction is applied together with PBE in all atomic-structure optimizations in this work \cite{Tkatchenko2009}. Charge transfer in NO$_3$-doped CNT systems is investigated also with GPAW by applying the Perdew-Zunger SIC to DFT calculations \cite{Perdew1981,Gudmundsdottir2015}. In this case, the exchange-correlation functional is chosen to be PW91 \cite{Perdew1992}. The PW91 energy functional is corrected by subtracting a SIC term that is scaled by a factor of $1/2$ (for details, see Ref.~\cite{Gudmundsdottir2015}).

\begin{figure}[!ht]
\centering
\includegraphics[scale=0.35]{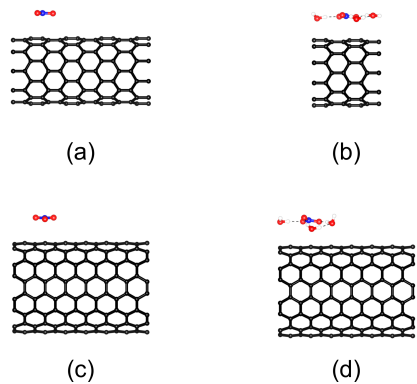}
\caption{Computational unit cells of (a) a semiconducting (10,0) zigzag CNT with one NO$_3$ molecule, (b) a NO$_3$-doped (10,0) CNT with three H$_2$O molecules, (c) a metallic (8,8) armchair CNT with one NO$_3$ molecule, and (d) an (8,8) CNT with one NO$_3$ molecule coordinated to three H$_2$O molecules. The computational (10,0) unit cells comprise either two or four primitive unit cells. The computational (8,8) unit cell contains seven primitive unit cells.\label{fig:unit_cells}}
\end{figure}

The effect of nitric-acid doping is considered in infinitely long NO$_3$-doped semiconducting (10,0) and metallic (8,8) CNTs by computing band structures for the computational unit cells depicted in Figs.~\ref{fig:unit_cells}(a)--(d). Our basic doped CNT system has only one NO$_3$ molecule in the computational unit cell of either a (10,0) or an (8,8) CNT. Higher doping concentrations are investigated by increasing the number of NO$_3$ molecules in the basic system. The concentration of NO$_3$ molecules in the investigated systems varies between $0.06$ and $0.47$ molecules per \r{A}. Further, the influence of water on the doping efficiency is studied using the systems shown in Figs.~\ref{fig:unit_cells}(b) and \ref{fig:unit_cells}(d), where a NO$_3$ molecule lying on the CNT is coordinated to three H$_2$O molecules.

The geometries of the primitive unit cells of (10,0) and (8,8) CNTs are first optimized with the FHI-aims code. The initial positions of the carbon atoms are computed by using the TubeGen 3.4 nanotube generator \cite{Frey2009} and the whole primitive unit cell including the lattice vectors is relaxed. In addition, the structures of the molecules added on the CNT are relaxed separately. Thereafter, a NO$_3$-doped CNT system is constructed with the help of the relaxed molecule and primitive CNT unit cells. The lengths of the computational unit cells of the (8,8) and (10,0) systems are $17.3$ and $17.1$ \r{A}, respectively. This means that the computational unit cells of the (8,8) and (10,0) systems comprise seven and four primitive CNT unit cells, respectively. In the case of the doped (10,0) system with H$_2$O molecules, the computational unit cell is shorter, around $8.5$ \r{A}. This system has to have a smaller number of atoms than in the other CNT structures so that Perdew-Zunger SIC calculations can be carried out. Periodic boundary conditions are used and the distance between the axes of the CNTs in the neighboring unit cells is $30.0$ \r{A}. The relaxation is performed with PBE and stopped when all the force components have decreased below $10^{-2}$ eV/\r{A}. A $1 \times 1 \times 27$ k-point grid is used when the computational unit cells of the doped systems are relaxed. The calculations of semiconducting (10,0) CNTs with NO$_3$ molecules can include spin since in some cases it gives a ground state with a lower energy than a non-spin-polarized system.

The binding energy $E_b$ for a NO$_3$ molecule on a CNT can be expressed as
\begin{equation}
E_b = E_\textrm{T}[\textrm{CNT-NO}_3] - E_\textrm{T}[\textrm{CNT}] - E_\textrm{T}[\textrm{NO}_3],
\label{eq:b_energy}
\end{equation} 
where $E_\textrm{T}[\textrm{CNT-NO}_3]$, $E_\textrm{T}[\textrm{CNT}]$, and $E_\textrm{T}[\textrm{NO}_3]$ are the computational unit cell total energies of the CNT with one NO$_3$ molecule, the pristine CNT, and the isolated NO$_3$ molecule, respectively. If the binding energy is determined for a doped system with water molecules, the total energy of the NO$_3$ molecule in Eq.~(\ref{eq:b_energy}) has to be replaced by that of the molecular complex. 

\begin{figure*}[!htb]
\centering
\includegraphics[scale=0.75]{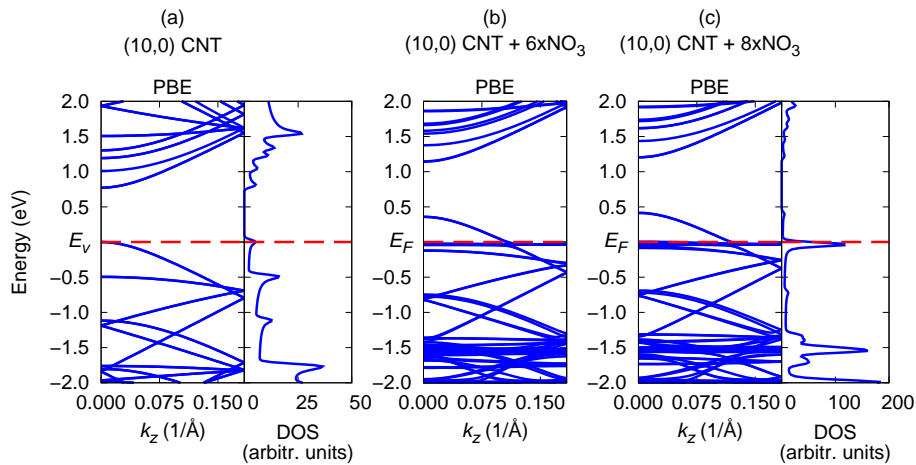}
\caption{Band structures of (a) a pristine (10,0) CNT and (b)--(c) NO$_3$-doped (10,0) CNTs. The DOSs for the systems studied are also shown in (a) and (c). The DOS peaks in (a) are van Hove singularities and the additional high DOS peaks in (c) are due to localized molecular states. The red dashed line represents the valence band maximum $E_v$ of the pristine (10,0) CNT or the Fermi level of the NO$_3$-doped (10,0) CNT.\label{fig:bands_sc_01}}
\end{figure*}

Electron transport in infinitely long homogeneously doped CNTs and in junctions of CNTs at the zero-bias limit is investigated using the transport module of the FHI-aims code. Two- or four-terminal transport systems can be divided into the central scattering region and semi-infinite leads. First, the electronic structure of the transport system relaxed with PBE and including the van der Waals correction has to be determined. Then one computes the electronic transmission function by solving the Green's function consistent with the PBE exchange-correlation potential for the CNT system \cite{Ferrer2014}. A significant challenge is aligning the energy levels of the semi-infinite leads and their counterparts in the computational transport unit cell. This alignment is discussed in Refs.~\cite{Ketolainen2017} and \cite{Havu2011}. The k-point grid used for finding the electronic structure in the transport calculations is $1 \times 1 \times 27$ for individual metallic (8,8) CNTs. A sparser grid is used for the transport calculations of semiconducting CNTs but the calculations are converged despite the different grid. In the case of a CNT junction, the k-point mesh is $1 \times 3 \times 3$.

\section{Results and discussion\label{sec:results}}

\subsection{Semiconducting CNTs with NO$_3$ molecules\label{sec:semiconducting_nanotubes}}

Relaxation of the computational unit cell of a semiconducting (10,0) CNT with one NO$_3$ molecule is carried out first with spin. After the relaxation, the center of the NO$_3$ molecule (the N atom) lies above a carbon atom. Our results are consistent with those of Peng \emph{et al.}~\cite{Peng2004}. Their results have been computed using the local spin-density approximation (LSDA). In a more recent study by Kroes \emph{et al.}~\cite{Kroes2016}, the optimal geometry is different so that the NO$_3$ molecule has moved a bit from the top configuration.

The distance between the CNT and the NO$_3$ molecule $d_{\textrm{CNT-NO}_3}$ is defined as a distance between the nitrogen atom and the carbon atom to which the NO$_3$ molecule binds. With the van der Waals correction, the distance between the NO$_3$ molecule and the (10,0) CNT is $3.09$ \r{A}. This value is significantly larger than the distance $2.87$ \r{A} obtained in an LSDA calculation. The overbinding of the LSDA is a known issue and is largely due to the rapid (exponential) decay of the LSDA potential. On the other hand, the CNT-NO$_3$ distance is close to the value $3.08$ \r{A} determined by Kroes \emph{et al.}~by using van der Waals corrected DFT \cite{Kroes2016}. The binding energy for one NO$_3$ molecule on a (10,0) CNT calculated using Eq.~(\ref{eq:b_energy}) is $-0.80$ eV in good agreement with the value found in Ref.~\cite{Kroes2016}. Semiconducting (10,0) CNTs doped with NO$_3$-H$_2$O complexes are also examined but the computational unit cell (see Fig.~\ref{fig:unit_cells}(b)) is shorter than that used for (10,0) CNTs with only NO$_3$ molecules. A detailed description of these systems is given in the appendix.

The band structures of pristine and NO$_3$-doped (10,0) CNTs are shown in Figs.~\ref{fig:bands_sc_01}(a)--(c) along with the densities of states (DOSs) for the pristine (Fig.~\ref{fig:bands_sc_01}(a)) and the most doped (Fig.~\ref{fig:bands_sc_01}(c)) CNTs. The system with either six or eight NO$_3$ molecules is investigated without relaxing the atomic structure except for the CNT-NO$_3$ distance because the molecules do not affect the CNT structure remarkably and the interaction between the molecules is small. The pristine (10,0) CNT has a direct band gap of $0.77$ eV at the $\Gamma$ point in the case of PBE as shown in Fig.~\ref{fig:bands_sc_01}(a).

A downshift of the Fermi level in the band structures of NO$_3$-doped (10,0) CNTs in Figs.~\ref{fig:bands_sc_01}(b) and \ref{fig:bands_sc_01}(c) is observed. In addition, there is a dispersionless molecular state just below the Fermi level in the band structures shown in Figs.~\ref{fig:bands_sc_01}(b) and \ref{fig:bands_sc_01}(c). A previous study of graphene doped with nitric acid has proposed that the work function of the half-filled orbital of the NO$_3$ molecule is greater than that of graphene and therefore the molecular state lies below the Fermi energy \cite{Lu2017}. Furthermore, the Fermi level in the highly doped system in Fig.~\ref{fig:bands_sc_01}(c) is pinned to a van Hove singularity and the flat molecular state. This is similar to our previous observation in the case of AuCl$_4$ doping and it is an important ingredient for establishing a robust enhanced intertube conduction \cite{Ketolainen2017}.

We estimate the charge transfer in the NO$_3$-doped systems by using the method described in Ref.~\cite{Ketolainen2017}. In this method, the crossing points of the Fermi level and the valence bands are first searched for. These points divide the bands into occupied and unoccupied parts. Thereafter, by measuring the width of the unoccupied regions and dividing this value by the maximum $k_z$ wave vector, an estimate for the magnitude of the charge transfer can be obtained when the orbital and spin degeneracies are taken into account.

The doping efficiency for one NO$_3$ molecule in the basic computational unit cell, i.e. the electron transfer from the CNT to NO$_3$ molecules, is $0.6$ electrons per molecule when the band structure is computed with PBE. Band structures for computational unit cells consisting of (10,0) CNTs with two NO$_3$ molecules are shown in the appendix and the charge transfer per one molecule in this system is $0.5$ electrons. The band structures for highly doped (10,0) CNTs in Figs.~\ref{fig:bands_sc_01}(b) and \ref{fig:bands_sc_01}(c) show significant downshifts of the Fermi level but the average doping efficiencies of the NO$_3$ molecules are smaller than those determined for the two-molecule system (see Table \ref{table:d_eff_sc}). Therefore, the doping efficiency decreases as the dopant density increases. In the case of two molecules in the computational unit cell, the total energy of the system is $0.15$ eV lower when spin is included in the calculation. No significant differences between the band structures of the NO$_3$-doped CNTs with and without spin-polarization are found but a clear (spin) splitting of the molecular state close to the Fermi level occurs in the spin-polarized case (see Fig.~\ref{fig:bs_two_mol}(b) in the appendix). Otherwise almost all of the bands in the $-2.0$ to $2.0$ eV range, relative to the Fermi level, overlap with each other in the spin-polarized band structure presented in Fig.~\ref{fig:bs_two_mol}(b). The doping efficiency determined for a computational unit cell with two NO$_3$ molecules taking the spin into account is approximately $0.4$ electrons per one molecule. 

A recent DFT study has revealed that the NO$_3$ molecule on a (10,0) CNT obtains a charge of $0.5$--$0.6$ e according to a Bader charge analysis \cite{Kroes2016}. The charge transfer found in the present work is rather close to that result. For comparison, we have determined the Hirshfeld charge for a single NO$_3$ molecule and obtained a value $0.4$ e. This is also in agreement with the Bader charge although slightly smaller than that given by the band structure analysis.

\begin{table}[!h]
\centering
\caption{Doping efficiency in NO$_3$-doped (10,0) CNTs per one molecule. The values are given per molecule as a function of the number of NO$_3$ molecules ($N_{\textrm{NO}_3}$) in the computational unit cell. The results have been computed with PBE.\label{table:d_eff_sc}}
\begin{tabular}{|c|c|c|}
\hline
\multirow{2}{*}{$N_{\textrm{NO}_3}$} & \multirow{2}{*}{Includes spin} & Doping efficiency \\
& & per NO$_3$ molecule (e) \\
\hline
$1$ & No & $0.6$ \\
$2$ & No & $0.5$ \\
$2$ & Yes & $0.4$ \\
$6$ & No & $0.4$ \\
$8$ & No & $0.3$ \\
\hline
\end{tabular}
\end{table}

\subsection{Metallic CNTs with NO$_3$ molecules\label{sec:metallic_nanotubes}}

\begin{figure*}[!htb]
\centering
\includegraphics[scale=0.75]{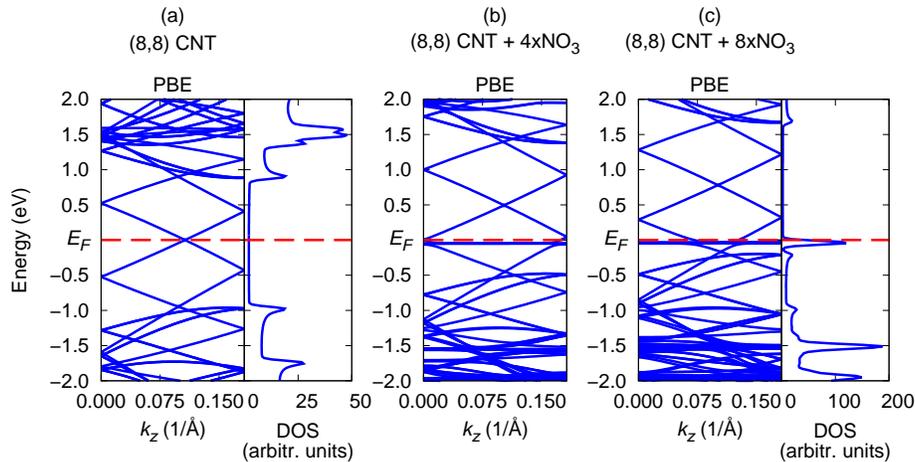}
\caption{Band structures of (a) a pristine (8,8) CNT and (b)--(c) NO$_3$-doped (8,8) CNTs. The DOSs for the systems studied are also shown in (a) and (c). The red dashed lines denote the Fermi level. The origin of the DOS peaks is the same as in Fig.~\ref{fig:bands_sc_01}.\label{fig:bands_metallic_01}}
\end{figure*}

The number of NO$_3$ molecules in the computational unit cell of a metallic (8,8) CNT is varied and the atomic structure of each doped system is optimized separately. After the relaxation, the nitrogen atom of the NO$_3$ molecule resides, as in the case of the semiconducting (10,0) CNT, on top of a carbon atom and the oxygen atoms are located near the centers of the carbon hexagons. The CNT-NO$_3$ distance in a NO$_3$-doped (8,8) CNT shown in Fig.~\ref{fig:unit_cells}(c) is $3.06$ \r{A}. The binding energy given by Eq.~(\ref{eq:b_energy}) for a NO$_3$ molecule on an (8,8) CNT is $-1.31$ eV, indicating a more stable structure than the (10,0) CNT, which has a binding energy of $-0.80$ eV. The trend that the binding energy is higher for metallic CNTs is in accordance with experimental findings for semiconducting and metallic CNTs \cite{Ruiz-Soria2014}. It is due to the stronger polarization of the electron gas in metallic CNTs. The CNT-molecule distance does not change significantly when the number of NO$_3$ molecules in the computational unit cell of the CNT is increased from one to eight. The values of the average CNT-molecule distances and other details are given in the appendix.

The band structures of a pristine (8,8) CNT and NO$_3$-doped (8,8) CNTs are displayed in Figs.~\ref{fig:bands_metallic_01}(a)--(c). An (8,8) CNT without doping has a band structure of a metallic system and the Fermi level is located at a point where two bands cross each other. Placing NO$_3$ molecules on top of an (8,8) CNT results in a relative downshift of the Fermi level as can be seen from Fig.~\ref{fig:bands_metallic_01}(b). This downshift can be regarded as a p-type doping effect since there is charge transfer from the CNT states to the molecular ones. In addition, similarly to the case of the NO$_3$-doped semiconducting (10,0) CNTs, there is a flat band very close to the Fermi level that is a molecular state. A similar state has been observed in another study \cite{Guerini2008} but it was found lower below the Fermi level compared with this work. The downshift of the Fermi level can be enhanced by increasing the number of molecules in the computational unit cell. The Fermi level, however, does not reach the first van Hove singularity although the molecular concentration is increased to the same level as that of the highly doped semiconducting (10,0) CNT. This can be seen in Fig.~\ref{fig:bands_metallic_01}(c). Thus, obtaining a stable doping effect via the Fermi level pinning at the van Hove singularity can be challenging \cite{Ketolainen2017}.

\begin{table}[!h]
\centering
\caption{Doping efficiency in NO$_3$-doped (8,8) CNTs. The values are given per molecule as a function of the number of NO$_3$ molecules ($N_{\textrm{NO}_3}$) in the computational unit cell and have been computed with PBE.\label{table:d_eff_m}}
\begin{tabular}{|c|c|}
\hline
\multirow{2}{*}{$N_{\textrm{NO}_3}$} & Doping efficiency \\
& per NO$_3$ molecule (e) \\
\hline
$1$ & $0.7$ \\
$2$ & $0.6$ \\
$3$ & $0.6$ \\
$4$ & $0.5$ \\
$5$ & $0.5$ \\
$8$ & $0.4$ \\
\hline
\end{tabular}
\end{table}

The values for the charge transfer in NO$_3$-doped (8,8) CNTs are given in Table \ref{table:d_eff_m}. The doping efficiency of one NO$_3$ molecule on an (8,8) CNT per computational unit cell is close to $0.7$ electrons. A significant finding is also that the doping efficiency of NO$_3$ molecules decreases when the molecular concentration on the CNT is increased. If the number of NO$_3$ molecules around the (8,8) CNT is doubled, the doping efficiency decreases approximately by $0.1$ electrons per molecule. In the case of a computational unit cell with eight NO$_3$ molecules, the value of the charge transfer has decreased to $0.4$ electrons per molecule.

\subsection{Doping efficiencies determined with HSE06 and SIC\label{sec:results_other}}

The spurious self-interaction error inherent to practical implementations of DFT, particularly at the GGA level of theory, makes charge transfer systems notoriously difficult to describe \cite{Ruzsinszky2006}. This is due to incomplete cancellation of the self-Coulomb term (i.e. an electron feels its own Coulomb repulsion) in the exchange-correlation potential. Electrons tend to delocalize in space in order to minimize the self-repulsion and as a result the observed charge transfer, or effective doping, is lowered. This can be remedied by including (partial) exact-exchange -- as in hybrid exchange-correlation functionals -- or by explicitly subtracting the self-Coulomb term from the total energy functional. Therefore, we compare our PBE results shown above with those obtained by using the HSE06 functional and the Perdew-Zunger SIC approach. Computing the band structure for a NO$_3$-doped (10,0) CNT with HSE06 shows an enhancement of the downshift of the Fermi level with respect to the PBE results given above. This indicates a higher doping efficiency. Moreover, the molecular levels move toward lower energies with respect to the CNT bands. Besides the FHI-aims code, the other code package GPAW is used to compute charge transfer in a short computational unit cell of a NO$_3$-doped (10,0) CNT using the SIC. Perdew-Zunger SIC calculations have been successful in describing challenging systems where the commonly used exchange-correlation functionals of DFT are in error. This includes a localized defect state in a crystal \cite{Gudmundsdottir2015}, and charged localized states of a molecule \cite{Cheng2016}. In the case of NO$_3$-doped (10,0) CNTs, ground state calculations including the Perdew-Zunger SIC are performed first. Thereafter, we carry out a Bader charge analysis \cite{Henkelman2006} and determine the partial charges of the NO$_3$ molecule. If there is only one NO$_3$ molecule in the short computational unit cell and the SIC is included in the calculation, the value of the charge transfer is $0.9$ electrons. Therefore, the SIC method results in clearly higher doping efficiency for the NO$_3$ molecule than the PBE calculations.

\begin{figure}[!htb]
\centering
\includegraphics[scale=0.75]{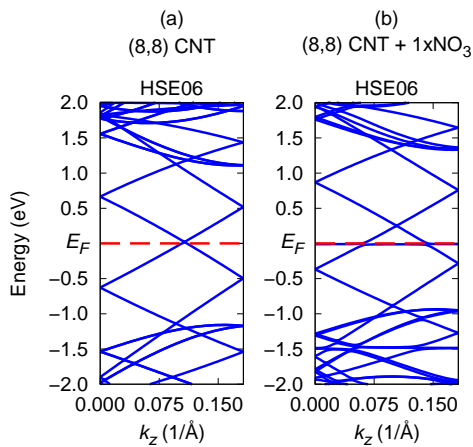}
\caption{Band structures computed with HSE06 for (a) a pristine (8,8) CNT and (b) an (8,8) CNT with one NO$_3$ molecule in the computational unit cell.\label{fig:bands_metallic_HSE06}}
\end{figure}

The band structures calculated with HSE06 for pristine and NO$_3$-doped (8,8) CNTs are presented in Figs.~\ref{fig:bands_metallic_HSE06}(a) and \ref{fig:bands_metallic_HSE06}(b). Compared to the PBE results in Fig.~\ref{fig:bands_metallic_01}(a), a shift of the bands below the Fermi level toward lower energies is observed in the band structure for a pristine (8,8) CNT (see Fig.~\ref{fig:bands_metallic_HSE06}(a)). Similarly to the case of a (10,0) CNT with one NO$_3$ molecule in the computational unit cell, the downshift of the Fermi level in Fig.~\ref{fig:bands_metallic_HSE06}(b) becomes more remarkable than in PBE calculations giving a higher doping efficiency when the calculation is carried out with the HSE06 functional. The doping efficiency increases to over $0.9$ electrons per molecule. The doping efficiencies of NO$_3$-doped semiconducting and metallic CNTs with different molecular concentrations are given in Table \ref{table:d_eff_other}. The Perdew-Zunger SIC value has been computed using a smaller computational unit cell than that of the HSE06 calculations. The functional in our transport calculations is PBE and the comparison between the PBE and HSE06 band structures indicates that our transport calculations will give qualitative results. However, they will predict phenomena taking place with doping and, thereby, they give insight into the electron transport mechanisms along individual CNTs and across CNT junctions in particular.

\begin{table}[!h]
\centering
\caption{Doping efficiency determined with HSE06 or the Perdew-Zunger SIC in (10,0) and (8,8) CNTs doped with one or two ($N_{\textrm{NO}_3}$) NO$_3$ molecules per computational unit cell.\label{table:d_eff_other}}
\begin{tabular}{|c|c|c|c|}
\hline
\multirow{2}{*}{Chirality} & \multirow{2}{*}{$N_{\textrm{NO}_3}$} & \multirow{2}{*}{Functional} & Doping efficiency \\
& & & per NO$_3$ molecule \\
\hline
(10,0) & $1$ & HSE06 & $0.8$ \\
(10,0) & $1$ & PW91 (SIC) & $0.9$ \\
(8,8) & $1$ & HSE06 & $0.9$ \\
(8,8) & $2$ & HSE06 & $0.8$ \\
\hline
\end{tabular}
\end{table}

\subsection{Carbon nanotubes with NO$_3$-H$_2$O complexes\label{sec:nanotubes_complex}}

\begin{figure}[!b]
\centering
\includegraphics[scale=0.75]{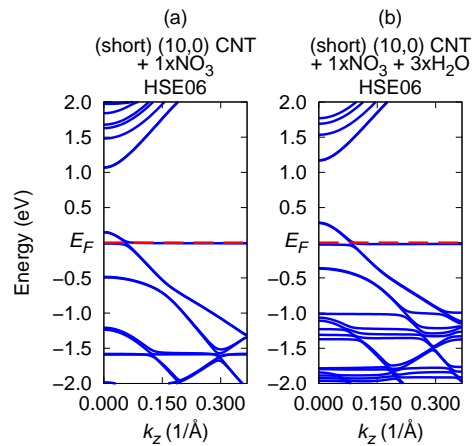}
\caption{Band structures computed with HSE06 for the short computational unit cells (see Fig.~\ref{fig:unit_cells}(b)) of NO$_3$-doped (10,0) CNTs (a) without and (b) with three H$_2$O molecules. The Fermi level is denoted by a red dashed line.\label{fig:bs_sc_water}}
\end{figure}

\begin{figure*}[!htb]
\centering
\includegraphics[scale=0.75]{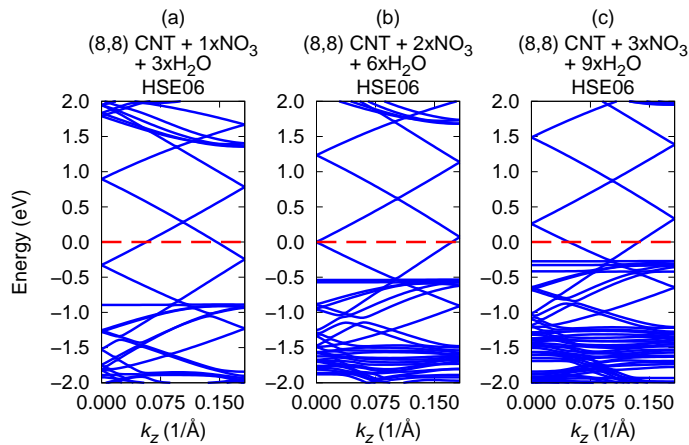}
\caption{Band structures computed with HSE06 for (8,8) CNTs with NO$_3$-H$_2$O molecular complexes. The number of molecular complexes on the CNT is (a) one, (b) two, or (c) three per computational unit cell. The Fermi level is marked with a red dashed line.\label{fig:bands_m_water}}
\end{figure*}

The charge transfer (or doping efficiency) can be enhanced by water molecules which coordinate to anionic species in molecular systems \cite{Jonsson2011}. Due to strong dielectric screening, water molecules stabilize charged atoms or molecules, facilitating a more chemically robust charge transfer, in the sense that a unit, or near unit, of charge is transferred between the donor and acceptor. Inspired by this observation, we have studied NO$_3$-H$_2$O complexes on semiconducting (10,0) and metallic (8,8) CNTs. After structure optimizations with the PBE functional and the van der Waals correction, band structures are calculated using the HSE06 functional.

For the (10,0) CNT, the calculations are performed with the short computational unit cell depicted in Fig.~\ref{fig:unit_cells}(b) in order to facilitate a comparison with Perdew-Zunger SIC results. The HSE06 band structures for one NO$_3$ molecule without or with three H$_2$O molecules in the computational unit cell are presented in Fig.~\ref{fig:bs_sc_water}. The comparison shows that water is capable of enhancing the charge transfer from the CNT. Without water, the charge transfer from the CNT to the NO$_3$ molecule is $0.6$ electrons, which is slightly larger than the PBE result of $0.5$ electrons for the same dopant concentration (see Table \ref{table:d_eff_sc}). With the three coordinated H$_2$O molecules, the doping efficiency is $0.9$ electrons per NO$_3$-H$_2$O complex. Performing the same calculation with the Perdew-Zunger SIC method gives a doping efficiency of $1.1$ electrons. Therefore, both the Perdew-Zunger SIC and HSE06 calculations predict a practically complete charge transfer of one electron from the CNT to the NO$_3$-H$_2$O complex.

In the case of the metallic (8,8) CNT, we have examined systems of several NO$_3$-H$_2$O complexes per computational unit cell. The HSE06 band structures for one, two, and three NO$_3$ molecules each coordinated to three H$_2$O molecules in the computational unit cell (see Fig.~\ref{fig:unit_cells}(d)) are presented in Fig.~\ref{fig:bands_m_water}. The remarkable qualitative effect of H$_2$O molecules is the pushing of molecular states clearly below the Fermi level as can be seen by comparing figures. As a result, the calculated doping efficiencies indicate in all the cases studied a complete charge transfer of one electron per NO$_3$-H$_2$O complex. This is in contrast to the efficiencies of $0.9$ and $0.8$ electrons per bare NO$_3$ molecule for one and two molecules in the computational unit cell, respectively (see Table \ref{table:d_eff_other}) since the doping efficiency decreases as the number of molecules in the system is increased. Figure \ref{fig:bands_m_water} shows also the increase of dispersionless molecular states in number due to coordinated H$_2$O molecules. Moreover, when the concentration of the molecular complexes increases, the splitting of the molecular states originating from the interactions between the large complexes enhances. At the same time, also part of the dispersive CNT states are split indicating some hybridization between the states of the CNT and the NO$_3$-H$_2$O molecular complexes. In conclusion, the H$_2$O molecules enhance the doping efficiency, but the increased interactions between the molecular and CNT states will increase the electron scattering and are harmful for the electron transport.

\subsection{Electron transport in doped individual (10,0) and (8,8) CNTs\label{sec:transport_individual}}

\begin{figure*}[!htb]
\centering
\includegraphics[scale=0.75]{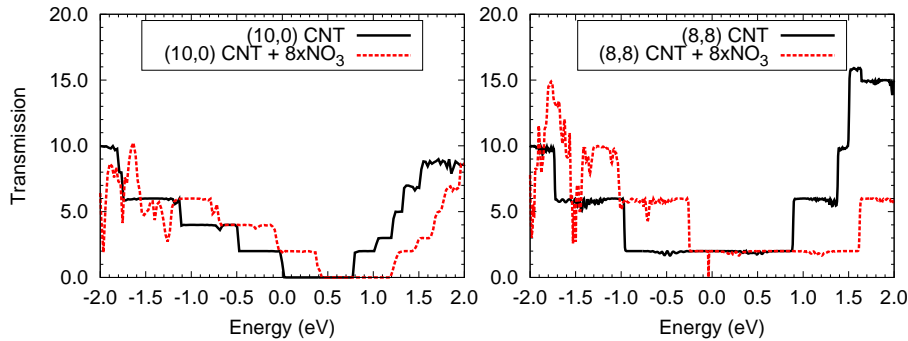}
\caption{Electronic transmission functions for individual pristine (black solid lines) and NO$_3$-doped (red dashed lines) CNTs. The systems in the left panel are a pristine semiconducting (10,0) CNT and a (10,0) CNT with eight NO$_3$ molecules in the computational unit cell. The right panel presents the transmission functions for a pristine metallic (8,8) CNT and for a metallic (8,8) CNT doped with eight NO$_3$ molecules in the computational unit cell. The energy zero denotes the top of the valence band of the pristine semiconducting (10,0) CNT or the Fermi level of the metallic systems.\label{fig:transport_single}}
\end{figure*}

Studying the conductivity of single NO$_3$-doped CNTs is done by constructing a two-terminal transport system by using the computational unit cells depicted in Figs.~\ref{fig:unit_cells}(a)--(d). Then the system contains three similar computational unit cells of which the first and third ones belong to the leads that are also doped. The middle cell forms the scattering region. The electronic transmission functions for individual semiconducting (10,0) and metallic (8,8) CNTs are shown in Fig.~\ref{fig:transport_single}. In addition to the transmission functions for pristine CNTs, the transport curves for NO$_3$-doped CNTs are presented in Fig.~\ref{fig:transport_single}.

In the case of a pristine (10,0) CNT, the transmission function possesses several steps and has a gap indicating semiconducting behavior. In contrast, there is no gap in the transmission function for the pristine (8,8) CNT. Therefore, this system is metallic. Corresponding to the changes in the band structures, in the case of both CNTs the transmission curves are shifted upwards in energy when the number of NO$_3$ molecules on the CNT is increased. Importantly, in the case of the doped semiconducting CNT the transmission at the Fermi level is remarkable, i.e., the doped CNT is metallic. Interestingly, a significant dip appears near the Fermi level in the transmission function for the NO$_3$-doped (8,8) CNT. The position of the dip corresponds to the energy of the molecular state in Figs.~\ref{fig:bands_metallic_01}(b)-(c). The decrease in the electronic transmission function at the Fermi energy can be attributed to Fano antiresonances in nanostructures \cite{Lambert2015}, effects found also, e.g., in graphene nanoribbon systems \cite{Saloriutta2011}. A decrease in the conductance has been observed in a computational study where a metallic (8,8) CNT with a NO$_2$ molecule has been placed between two Au (111) electrodes \cite{Sivasathya2014} so that our finding of the lowering of the transport in the NO$_3$ doping process is in agreement with the previous work.

\subsection{Junctions of doped CNTs\label{sec:junctions}}

The computational unit cells of doped CNTs investigated in the previous sections can also be used to form CNT junctions. In particular, we consider junctions of two perpendicular doped semiconducting (10,0) or metallic (8,8) CNTs. These junctions consist of CNTs with eight NO$_3$ molecules in the computational unit cell. Further, another junction between a NO$_3$-doped metallic (8,8) and a NO$_3$-doped semiconducting (10,0) CNT is investigated. In this junction, the numbers of NO$_3$ molecules in the computational unit cells of (8,8) and (10,0) CNTs are eight and six, respectively.

The distance between the doped CNTs is optimized before performing the transport calculation. When the van der Waals correction is taken into account, the distance between the (10,0) CNTs with NO$_3$ molecules is $3.21$ \r{A}. Thus, the distance is larger than $2.5 - 2.6$ \r{A} obtained in similar calculations for a junction of (8,8) CNTs without doping \cite{Havu2011,Ketolainen2015}, reflecting the effect of electron charge transfer from the CNTs to the molecules leaving the CNTs slightly electron deficient and resulting in Coulomb repulsion between the CNTs. The distance between the CNTs of the two other junctions are also optimized with the van der Waals correction. In the junction between two NO$_3$-doped (8,8) CNTs, the CNT-CNT distance is $3.16$ \r{A}. The distance in the third junction between a metallic (8,8) and a semiconducting (10,0) CNT is $3.11$ \r{A}. Thus, the doping increases the CNT-CNT distance but distances do not vary significantly when changing the chiralities of the CNTs.

\begin{figure}[!h]
\centering
\includegraphics[scale=0.75]{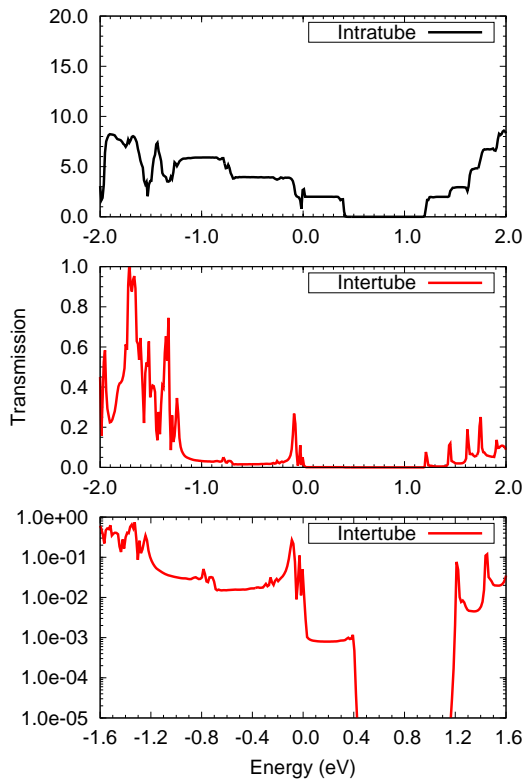}
\caption{Intratube (upper plot) and intertube (middle and lower plots) electronic transmission functions for a junction of two (10,0) CNTs with eight NO$_3$ molecules in the computational unit cell. The energy zero is at the Fermi level.\label{fig:transport_junction_01}}
\end{figure}

The intratube and intertube transmission functions for a junction of NO$_3$-doped (10,0) CNTs are presented in Fig.~\ref{fig:transport_junction_01}. The intratube transport in Fig.~\ref{fig:transport_junction_01} is rather good although the CNTs are doped with a large number of molecules. There are, however, small dips near the energy zero that are related to Fano antiresonances due to localized molecular states. The logarithmic plot in the bottom panel of Fig.~\ref{fig:transport_junction_01} shows an increase of about one decade in transmission when crossing a single van Hove singularity below the Fermi level. A similar increase can be seen in the transmission between two pristine (10,0) CNTs over the same singularity (published in Fig.~$6$ in our previous work \cite{Ketolainen2017}). This increase is clearly larger than the corresponding increase in DOS reflecting the fact that also changes in the wavefunctions and not only the number of states available affect the intertube transmission.

Besides the change in the average transmission over the Fermi level, the intertube transmission function in Fig.~\ref{fig:transport_junction_01} increases strongly just at the Fermi level, where there is a van Hove singularity and a molecular state. Thus, the increased DOS of the CNTs enhances the electron tunneling between the CNTs. Moreover, when comparing the intertube transmission with that between two pristine (10,0) CNTs (Fig.~$6$ in Ref.~\cite{Ketolainen2017}) we notice a clear broadening of the van Hove singularity derived peak just below the Fermi level. This means that the CNT and molecular levels are hybridized increasing the extent of the wavefunctions and the ensuing electron transmission. Because the Fermi level can be pinned to regions of high DOS and because the electric current is determined by integrating the transmission function over a Fermi-level centered bias window the total current through the junction will be increased. We should also note that for this reason the total current will increase due to doping although the intertube distance increases.

\begin{figure}[!h]
\centering
\includegraphics[scale=0.75]{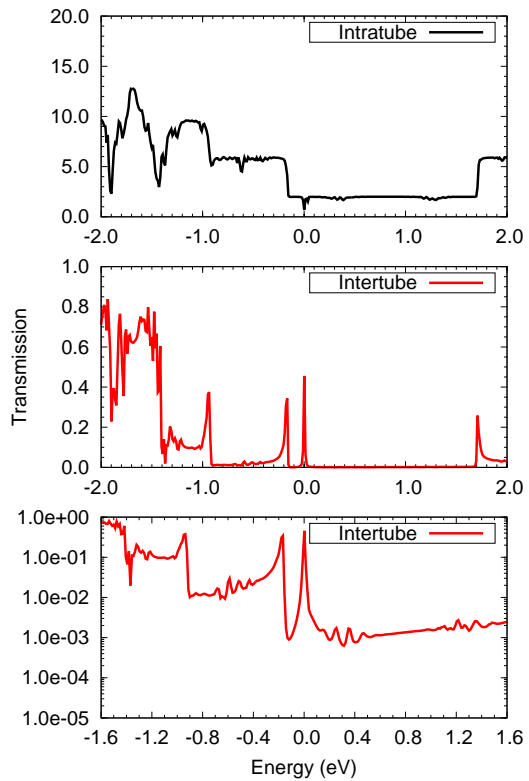}
\caption{Intratube (upper plot) and intertube (middle and lower plots) electronic transmission functions for a junction of two (8,8) CNTs with eight NO$_3$ molecules in the computational unit cell. The energy zero is at the Fermi level.\label{fig:transport_junction_02}}
\end{figure}

The results from the four-terminal transport calculations for a junction of NO$_3$-doped (8,8) CNTs are displayed in Fig.~\ref{fig:transport_junction_02}. In this case, electron transport along individual CNTs (see the upper plot in Fig.~\ref{fig:transport_junction_02}) remains almost as good as in pristine metallic (8,8) CNTs. The values of the intertube transmission function increase over the van Hove singularity nearly by one decade and the singularity causes a sharp peak. There is also a sharp and high peak close to the Fermi energy. This peak, which is due to tunneling through localized molecular states, is higher than a similar one found in the intertube function in Fig.~\ref{fig:transport_junction_01}. When we compare the intertube transmission in Fig.~\ref{fig:transport_junction_02} with that between two pristine (8,8) CNTs (see Fig.~$3$b in our previous article \cite{Ketolainen2015}), we notice that the transmission peaks corresponding to van Hove singularities below the Fermi level have similar widths. These observations about the peak shapes mean that the hybridization between the CNT and molecular states is minimal, the extents of the wavefunctions do not increase, and a doping-induced increase in the total intertube current is expected to remain lower than for the junction of two doped semiconducting (10,0) CNTs.

\begin{figure}[!h]
\centering
\includegraphics[scale=0.35]{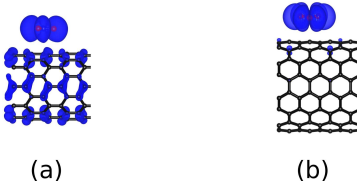}
\caption{Charge density of the highest occupied eigenstate of (a) a (10,0) CNT and (b) an (8,8) CNT with one NO$_3$ molecule in the computational unit cell. The value of the electron density of the isosurface is the same for both systems.\label{fig:dens_CNT_NO3}}
\end{figure}

The enhancement of the electron transport through the junction of NO$_3$-doped (10,0) CNTs relative to that of NO$_3$-doped (8,8) CNTs follows from enhanced hybridization of the CNT and molecular states as shown by the charge densities of the highest occupied eigenstates of the CNT-NO$_3$ systems (see Figs.~\ref{fig:dens_CNT_NO3}(a) and \ref{fig:dens_CNT_NO3}(b)). The spreading of the wavefunctions and the hybridization between the CNT and molecular orbitals is remarkably larger in the NO$_3$-doped (10,0) CNTs than in the NO$_3$-doped (8,8) CNTs, which leads to spreading of the eigenstates in energy and an ensuing increase in the intertube transmission over wide energy regions between broadened transmission peaks at van Hove singularities.

\begin{figure}[!h]
\centering
\includegraphics[scale=0.75]{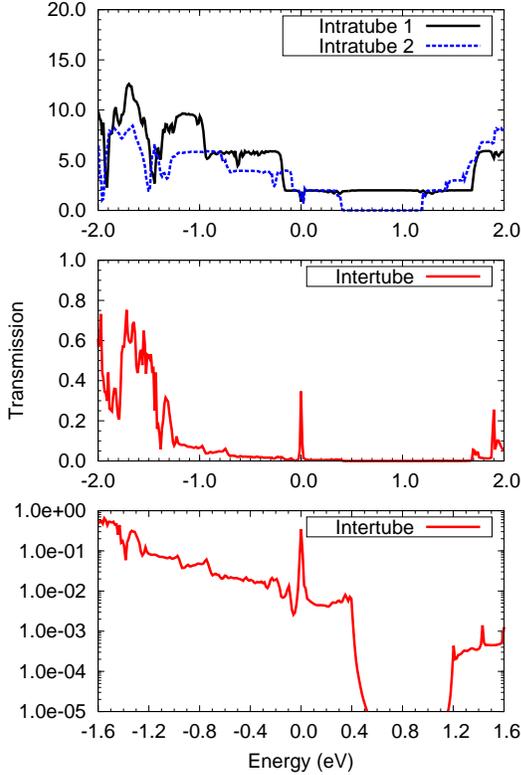}
\caption{Intratube (upper plot) and intertube (middle and lower plots) electronic transmission functions for a junction of a NO$_3$-doped (8,8) and a NO$_3$-doped (10,0) CNT. The numbers of NO$_3$ molecules in the computational unit cells of the (8,8) and (10,0) CNTs are eight and six, respectively. The energy zero is located at the Fermi level.\label{fig:transport_junction_03}}
\end{figure}

Performing a four-terminal transport calculation for a junction consisting of a doped metallic (8,8) and a doped semiconducting (10,0) CNT shows that the intratube transmission curves (see the upper panel of Fig.~\ref{fig:transport_junction_03}) of this junction are similar to those presented in Figs.~\ref{fig:transport_junction_01} and \ref{fig:transport_junction_02}. The intertube transmission function in the two lowest plots of Fig.~\ref{fig:transport_junction_03} indicates that the transmission through the junction below the band gap of the (10,0) CNT is rather high in comparison with the symmetric junctions discussed above. Moreover, the molecular state at the Fermi level causes a very narrow transmission peak. However, clear steps and wider regions of higher transmission are absent until rather low energies. This is due to the fact that the regions of high DOS due to van Hove singularities and molecular states in the two CNTs do not match at the same energies. Therefore, the conductances of junctions between doped semiconducting and metallic CNTs are expected to be low.

Intra- and intertube transmission functions in Fig.~\ref{fig:transport_junction_01} resemble those computed for a similar CNT junction of two AuCl$_4$-doped (10,0) CNTs \cite{Ketolainen2017}. The intratube transmission function for the NO$_3$-doped system, however, is smoother, i.e. has less peaks per energy unit, than that of the CNT junction with AuCl$_4$ molecules. This indicates that there are fewer molecular states in the NO$_3$ molecules and they do not hybridize with those of the CNT so strongly causing less electron scattering than the AuCl$_4$ molecules on the same CNT. Correspondingly, the number of peaks in the vicinity of the Fermi level in the intertube transmission function for (10,0) CNTs with NO$_3$ molecules is smaller than in the same region in the intertube function for AuCl$_4$-doped (10,0) CNTs and doping with NO$_3$ is expected to be less efficient than doping with AuCl$_4$.

Increases in the conductances of the junctions of doped CNTs are related to the Fermi level crossing van Hove singularities and the high and wide peaks of the intertube transmission functions. Similarly to the junctions of AuCl$_4$-doped (10,0) CNTs \cite{Ketolainen2017}, the conductances of junctions of NO$_3$-doped semiconducting (10,0) or metallic (8,8) CNTs enhance due to doping when the Fermi level shifts to a region, where the DOS is high. The Fermi level can be pinned to the van Hove singularities or localized molecular states leading to a robust mechanism for the improvement of the conductivity of CNT networks \cite{Ketolainen2017}. The pinning of the Fermi level at regions of high DOSs can also increase the conductances between dissimilar CNTs such as in the case of semiconducting (10,0) and metallic (8,8) CNTs discussed above.

Using atomic force microscopy, Znidarsic \cite{Znidarsic2013} et al.~measured resistances of single-wall CNTs and their junctions forming CNT networks before and after nitric acid treatments. They analyzed the measured data for junction resistances as a function of the diameters of the two CNTs forming X- or Y-shape junctions and as a function of the angle between the CNTs. The smallest contact resistances found were $29$ k$\Omega$. The contact resistances decreased with increasing CNT radii and decreasing the contact angle. Y-junctions showed smaller resistances than X-junctions. Because the unity transmission corresponds to the conductance quantum $2\textrm{e}^2/\textrm{h}$, the junction resistance of $29$ k$\Omega$ would correspond to a transmission of about $0.5$. According to our results in Figs.~\ref{fig:transport_junction_01}, \ref{fig:transport_junction_02}, and \ref{fig:transport_junction_03}, these kinds of transmission values are not plausible for the rather small radii (10,0) and (8,8) CNTs forming a perpendicular junction outside the transmission singularity peak regions unless the Fermi level has crossed three van Hove singularities. According to Ref.~\cite{Znidarsic2013}, the nitrogen acid treatment decreased the average junction resistance by a factor of around three, which may signalize the lowering of the Fermi level in CNTs and also its pinning around the van Hove singularities as a function of increased doping.

In a recent experimental work, Tsebro at al.~\cite{Tsebro2016} measured, as a function of temperature, sheet resistances of CNT networks before and after iodine or CuCl doping. The CNT radii were rather large, around $2$ nm, and the dopants filled the CNTs. Tsebro et al.~analyzed their data in terms of intra- and interbundle contributions corresponding to phonon scattering and fluctuation-assisted tunneling, respectively. With the help of calculated phonon dispersion relations and electronic band structures, they were able to conclude that iodine and CuCl doping lower the Fermi levels by $0.6$ and $0.9$ eV, respectively. The latter figure means that the Fermi level has crossed in semiconducting (metallic) CNTs three (one) van Hove singularities (singularity). The sheet resistances of the CNT networks decreased almost by a decade in the CuCl doping and the effect did not decay indicating the the internal doping of the CNT is stable. Our results show that the crossing of the Fermi level below a van Hove singularity causes a decrease of the junction resistance, which is of the same order of magnitude. Related to this, we also note that Tsebro et al.~proposed that the main cause for the sheet resistance of the CNT network is the junction resistance and that doping mainly decreases this contribution.

\section{Conclusions\label{sec:conclusions}}

The influence of NO$_3$ molecules on the band structures and electronic transmission functions of semiconducting and metallic CNTs has been explored to understand the nitric-acid doping process used to improve the conductivity of CNT thin films. In the present work, (10,0) and (8,8) CNTs are used in our density functional modeling to represent semiconducting and metallic CNTs, respectively. Our results show that electrons are transferred from the CNT to the NO$_3$ molecule and this results in p-type doping of the CNT. The average doping efficiency per NO$_3$ molecule decreases when the number of molecules on the CNT is increased. However, the doping efficiency enhances considerably when the NO$_3$ molecules form complexes with H$_2$O molecules.

Electron transport calculations of individual NO$_3$-doped semiconducting CNTs reveal that doping converts them metallic although sharp dips in the transmission function due to scattering off localized molecular states appear. The transmission functions increase stepwise over the van Hove singularities of the CNT electronic structures. Reflecting qualitatively different band structures, shifting the Fermi level toward the next singularity requires less doping in the case of semiconducting CNTs in comparison with metallic CNTs. This would favor semiconducting CNTs in CNT networks.

The intertube transmission functions between two semiconducting or two metallic CNTs increase around one decade over a van Hove singularity. This increases directly the conductance of the CNT junction. Moreover, the pinning of the Fermi level at the van Hove singularity and partially filled molecular states with the hybridization of the molecular and CNT states there can further increase the junction conductance and the conductivity of the CNT network. The hybridization of the molecular and CNT states is stronger for semiconducting than metallic CNTs, which would also favor the former in CNT networks. The energy mismatch between the high DOS regions for the clearly dissimilar electronic structures of metallic and semiconducting CNTs hinders to enhance the conductance of their junctions.

\appendix*
\section*{Appendix}
\setcounter{section}{1}

The geometries of metallic (8,8) and semiconducting (10,0) CNTs with NO$_3$ molecules are optimized using the FHI-aims code package. The number of NO$_3$ molecules on the CNTs is varied and the atomic structures of individual doped CNTs are relaxed. The relaxation is started with the light settings for the basis functions. After completing the relaxation with the light settings, the relaxation is continued with the tight settings. The relaxation calculations are stopped when the maximum force component falls below the tolerance values as explained in the article text. The optimal geometries of doped metallic and doped semiconducting CNTs are considered in the following sections. Some systems also have H$_2$O molecules near the NO$_3$ molecule. Moreover, this material contains a few figures of band structures for NO$_3$-doped CNTs and for these systems with H$_2$O molecules. The k-point grids of the band structure calculations for the PBE and HSE06 functionals are $1 \times 1 \times 27$ and $1 \times 1 \times 15$, respectively.

Optimizing the geometry of the computational unit cell of an (8,8) CNT with one NO$_3$ molecule shows that the distance between the NO$_3$ molecule and the CNT decreases by $0.16$ \r{A} when the Tkatchenko-Scheffler van der Waals correction is taken into account. The CNT-molecule distances for NO$_3$-doped (8,8) CNTs are listed in Table \ref{table:dist_met}. The distances are averages over all CNT-molecule distances in the system and have been calculated with the van der Waals correction.

\begin{table}[!h]
\centering
\caption{Average CNT-molecule distances for NO$_3$-doped (8,8) CNTs when taking the van der Waals correction into account.\label{table:dist_met}}
\begin{tabular}{|c|c|}
\hline
$N_{\textrm{NO}_3}$ & $d_{\textrm{CNT-NO}_3}$ (\r{A}) \\
\hline
$1$ & $3.06$ \\
$2$ & $3.05$ \\
$3$ & $3.07$ \\
$4$ & $3.07$ \\
$5$ & $3.07$ \\
$8$ & $3.07$ \\
\hline
\end{tabular}
\end{table}

The average CNT-molecule distance does not depend on the number of NO$_3$ molecules remarkably. Only a small increase occurs in the distances when the dopant concentration is increased. This change, however, is negligible.

Placing water molecules around the NO$_3$ dopant affects the CNT-molecule distance a little. The distance between the (8,8) CNT and the NO$_3$ molecule increases to $3.29$ \r{A} if the calculation includes the van der Waals correction. Neglecting this correction leads to an even larger CNT-molecule distance, $3.55$ \r{A}.

\begin{figure}[!htb]
\centering
\includegraphics[scale=0.75]{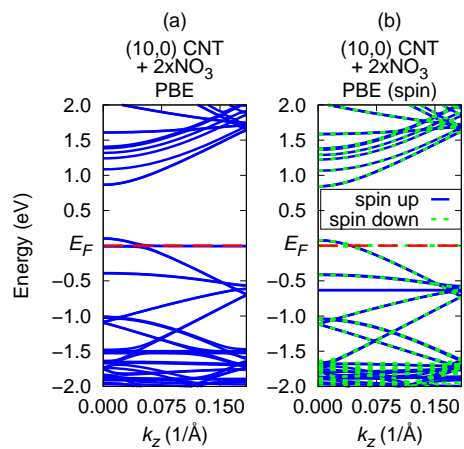}
\caption{Band structures computed (a) without and (b) with spin for the computational unit cells of (10,0) CNTs doped with two NO$_3$ molecules.\label{fig:bs_two_mol}}
\end{figure}

Optimization of the geometries of (10,0) CNTs doped with NO$_3$ molecules is performed with spin when there is only one molecule on a (10,0) CNT. The (10,0) CNTs with two NO$_3$ molecules are relaxed both with and without spin. In the case of (10,0) CNTs with NO$_3$-H$_2$O complexes, spin is omitted in the relaxation calculations. After optimizing the short computational unit cell of a NO$_3$-doped (10,0) CNT without water, the distance between the CNT and the molecule $d_{\textrm{CNT-NO}_3}$ is $3.05$ \r{A}. Thus, the molecule moves closer to the CNT compared with the geometry of a larger computational unit cell. The movement probably occurs because the charge of the CNT becomes larger on average when the molecular concentration is increased. As a result, the electrostatic interaction in the system pulls the molecules closer to the CNT. Adding H$_2$O molecules to the NO$_3$-doped (10,0) CNT system makes the molecular complex move further from the CNT surface. The distance $d_{\textrm{CNT-NO}_3}$ is $3.16$ \r{A} and the alignment of the molecular system changes because of the water molecules.

The downshift of the Fermi level occurs in the band structure for a (10,0) CNT with one NO$_3$ molecule in the computational unit cell. A larger downshift can be achieved by placing another molecule on the CNT in addition to the first one (see Fig.~\ref{fig:bs_two_mol}(a)). Carrying out a spin-polarized band structure calculation results in splitting of the energy levels near the Fermi energy region as shown in Fig.~\ref{fig:bs_two_mol}(b). The average doping efficiencies for the NO$_3$ molecules in this system can be found in the article text.

\begin{acknowledgments}
The computational resources provided by CSC - IT Center for Science Ltd.~have been of great importance. We would like to thank Prof.~Esko I.~Kauppinen for inspiring discussions. This work was supported by the Academy of Finland through its Centres of Excellence Programme (2012-2017) under Project No.~251748. In addition, the work was funded by the AEF Aalto Energy Efficiency research programme.
\end{acknowledgments}


\bibliography{ref_list_02}

\end{document}